\documentclass[reprint, amsmath,amssymb,aps,twocolumn, prb,superscriptaddress]{revtex4-1}
\pdfoutput=1
\usepackage{amsfonts,amssymb}
\usepackage[subnum]{cases}
\usepackage{mathrsfs}
\usepackage{amsmath}
\usepackage[none]{hyphenat}
\usepackage{hyperref}    
\usepackage{bm}          
\usepackage{natbib}
\usepackage{soul} 
\usepackage{color}
\usepackage{longtable}
\usepackage{ textcomp }
\usepackage{verbatim}
\usepackage{fancyhdr} 
\usepackage{lastpage} 
\usepackage{extramarks} 
\usepackage{graphicx} 
\usepackage{lipsum} 
\usepackage{ amssymb }
\usepackage{slashed}
\usepackage{tikz}
\usepackage{comment}
\usepackage{natbib}
\usepackage[caption=false]{subfig}

\usepackage{amsthm} 

\let\baraccent=\= 
\renewcommand{\=}[1]{\stackrel{#1}{=}} 

\theoremstyle{definition}

\theoremstyle{remark}

\begin{document}
\title{Electron Accumulation Layer in Ultrastrong Magnetic Field}	
\author{M. Sammon}
\email{sammo017@umn.edu}
\affiliation{Fine Theoretical Physics Institute, University of Minnesota, Minneapolis, MN 55455, USA}
\author{Han Fu}
\affiliation{Fine Theoretical Physics Institute, University of Minnesota, Minneapolis, MN 55455, USA}
\author{B. I. Shklovskii}
\affiliation{Fine Theoretical Physics Institute, University of Minnesota, Minneapolis, MN 55455, USA}
\date{\today}
\begin{abstract}
	When a three-dimensional electron gas is subjected to a very strong magnetic field, it can reach a quasi-one-dimensional state in which all electrons occupy the lowest Landau level. This state is referred to as the extreme quantum limit (EQL) and has been studied in the physics of pulsars and bulk semiconductors. Here we present a theory of the EQL phase in electron accumulation layers created by an external electric field $E$ at the surface of a semiconductor with a large Bohr radius such as InSb, PbTe, SrTiO$_3$ (STO), and particularly in the LaAlO$_3$/SrTiO$_3$ (LAO/STO) heterostructure. The phase diagram of the electron gas in the plane of the magnetic field strength and the electron surface concentration is found for different orientations of the magnetic field. We find that in addition to the quasi-classical metallic phase (M), there is a metallic EQL phase, as well as an insulating Wigner crystal state (WC). Within the EQL phase, the Thomas-Fermi approximation is used to find the electron density and the electrostatic potential profiles of the accumulation layer.  Additionally, the quantum capacitance for each phase is calculated as a tool for experimental study of these phase diagrams.
\end{abstract}	

\maketitle

\section{Introduction}
When a degenerate electron gas at low temperature $T$ is subjected to an ultrastrong magnetic field, its properties undergo dramatic changes. Particularly, when the external field $B$ is so strong such that 
\begin{equation}
\hbar \omega_c\gg E_F\gg k_BT,\frac{e^2n^{1/3}}{\kappa}, \label{eq:EQLdef}
\end{equation}
 the cyclotron energy becomes the dominant energy scale in the system. Here $\omega_c=eB/m^*c$ is the cyclotron frequency, $m^*$ is the effective mass, $E_F\approx\hbar^2n^{2/3}/2m$ is the Fermi energy at $B=0$, $k_BT$ is the thermal energy, $\kappa$ is the dielectric constant, and $n$ is the three dimensional concentration of electrons. When Eq. (\ref{eq:EQLdef}) is satisfied, we say that the system is in the ``extreme quantum limit" (EQL).

 Under the influence of a magnetic field $\boldsymbol{B}$, the kinetic energy of electrons in the direction perpendicular to $\boldsymbol{B}$ is quantized into Landau levels. In the EQL, the gap between adjacent levels becomes very large and electrons occupy the lowest Landau level only. As a result, the energy of the electron gas depends only on the momentum in the direction parallel to $\boldsymbol{B}$, creating a quasi-one-dimensional state. It has been proposed that under such conditions, various instabilities such as charge density waves, spin density waves, or Wigner crystallization occur.\cite{CelliSDW,KaplanWigner,Kleppmann,HalperinInstability}
 
  \begin{figure}[h!]
  	\includegraphics[width=7cm, height=5cm]{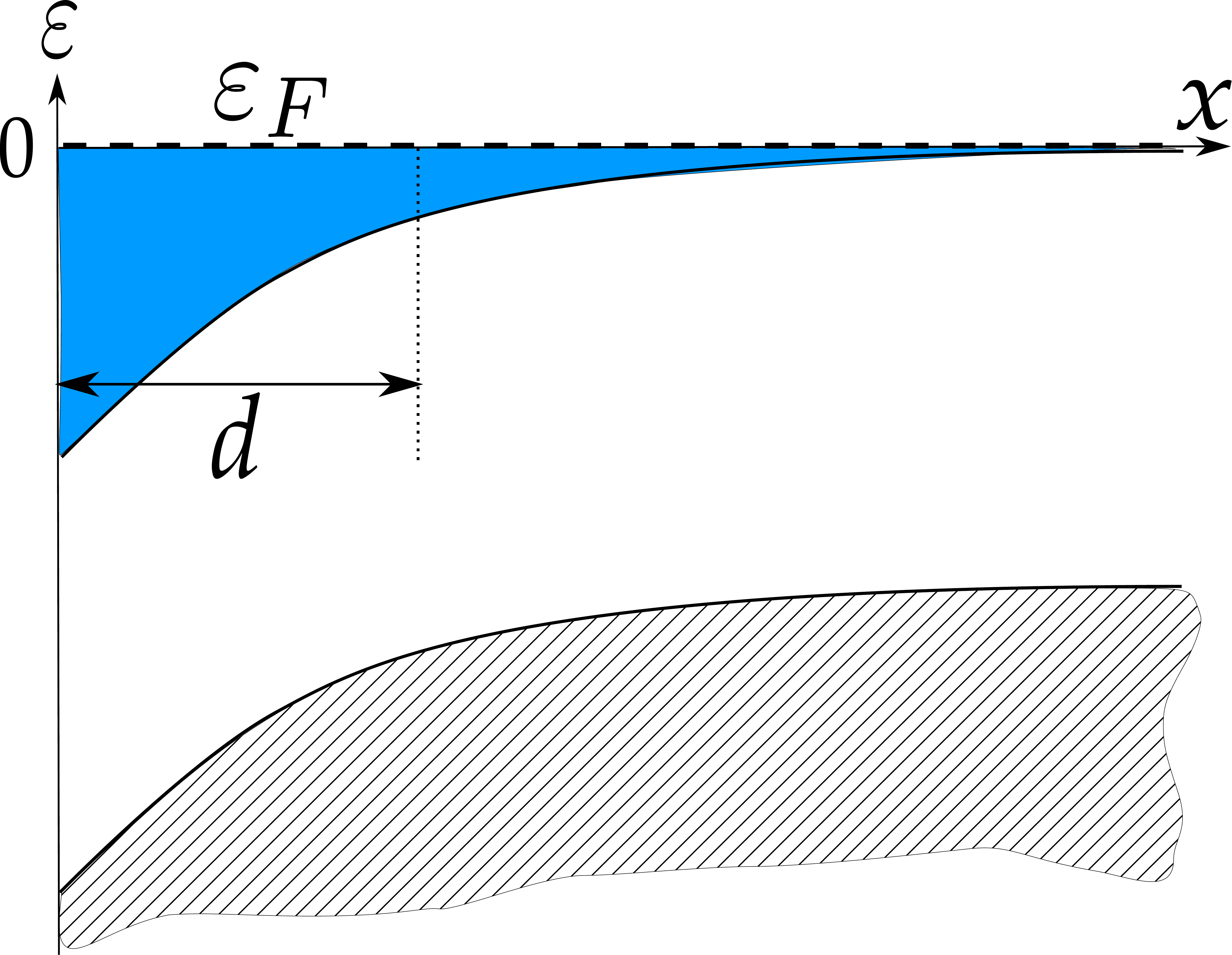}
  	\caption{(Color Online) Schematic energy diagram of an accumulation layer in a lightly doped n-type semiconductor, where $\varepsilon$ is the energy and $x$ measures the distance from the surface. Electrons (blue/dark grey layer) are attracted to the surface by an external electric field $E$, where they form the accumulation layer with a characteristic width $d$. In the bulk of the semiconductor, the Fermi level $\varepsilon_{F}$ lies near the bottom of the conduction band.}
  	\label{fig:accumulation layer}
  \end{figure}
  
What conditions are necessary to reach the EQL experimentally? From Eq. (\ref{eq:EQLdef}), it follows that in order for the gas to remain metallic, one must have $na^3\gg 1$. Here
 \begin{equation}
 	a=\kappa\hbar^2/m^*e^2\label{eq:bohr radius}
 \end{equation}
is the effective Bohr radius of the material. Additionally, the strong magnetic field condition, $\hbar\omega_c\gg E_F$, requires $n\lambda^3\ll 1$, where I have introduced $\lambda=\sqrt{\hbar c/eB}$ as the magnetic length. Combining $na^3\gg 1$ and $n\lambda^3\ll 1$, we find that in order to reach the EQL, we require $\lambda\ll a$. At $10$ T, $\lambda\approx10$ nm, and so we require materials in which $a\gg 10$ nm. 
 There are special materials such as InSb and  Hg$_{1-x}$Cd$_x$Te in which $a$ ranges between $60-120$ nm, so that the EQL is achievable at reasonable magnetic fields.\cite{Tokumoto,Shayegan,Murzin1d} In particular, bulk transport studies of InSb have found an experimental phase diagram that consists of a metal, EQL, and insulator phase.\cite{Murzin1d}

Another material in which the EQL may be reached is bulk SrTiO$_3$ (STO). STO is a semiconductor with a relatively heavy effective mass $m^*=1.5$ $m_e$,\cite{Lin} but a dielectric constant that becomes very large, $\kappa=2 \times 10^4$, at liquid helium temperatures.\cite{CowleyLattice, BarrettDielectric} As a result, the Bohr radius of STO becomes $a=700$ nm. This should create an ideal situation to study the EQL and several studies of the bulk magnetic properties of the material have been conducted,\cite{SonEpi,AllenSDH, KozukaEQL,Skinner}. Despite such effort, attempts to observe the EQL in bulk STO have not met much success, presumably due to disorder effects.\cite{Skinner}

Much attention has been devoted in recent years to LaAlO$_3$/SrTiO$_3$ (LAO/STO) heterostructures in which the ``polar catastrophe"\cite{OhtomoMobility, ThielTuning} creates an electric field that causes a high mobility electron gas to form at the interface. Recent magnetotransport studies of these structures have reported Integer Quantum Hall Effect steps in $\rho_{xy}$ that may be evidence of the gas approaching the EQL.\cite{TrierQHE, XieQHE, MatsubaraQHE}  

In this paper, we study the conditions under which one can observe the EQL in electron accumulation layers in semiconductors with a given dielectric constant $\kappa$ and Bohr radius $a$. Such an accumulation layer can be created in many ways. One example already mentioned is the polar catastrophe in LAO/STO heterostructures which creates an accumulation layer at the interface. Other common techniques include ionic liquid gating\cite{Ueno} of the semiconductor surface and $\delta$-doping by donors in the bulk of the sample.\cite{Jalan}  In all such cases the end result is an external electric field $\boldsymbol{E}$ along the direction perpendicular to the surface that causes electrons to accumulate near the surface (See Fig. \ref{fig:accumulation layer}). We can always relate $E$ to the surface concentration $N$ of electrons in the accumulation layer by 
\begin{equation}
E=\frac{4\pi eN}{\kappa}.\label{eq:Surface Charge}
\end{equation}
In our discussion below, all results are expressed through the surface concentration $N$ rather than the external field $E$. 

\begin{figure}[t]
	\includegraphics[width=7.29cm, height=6.8cm]{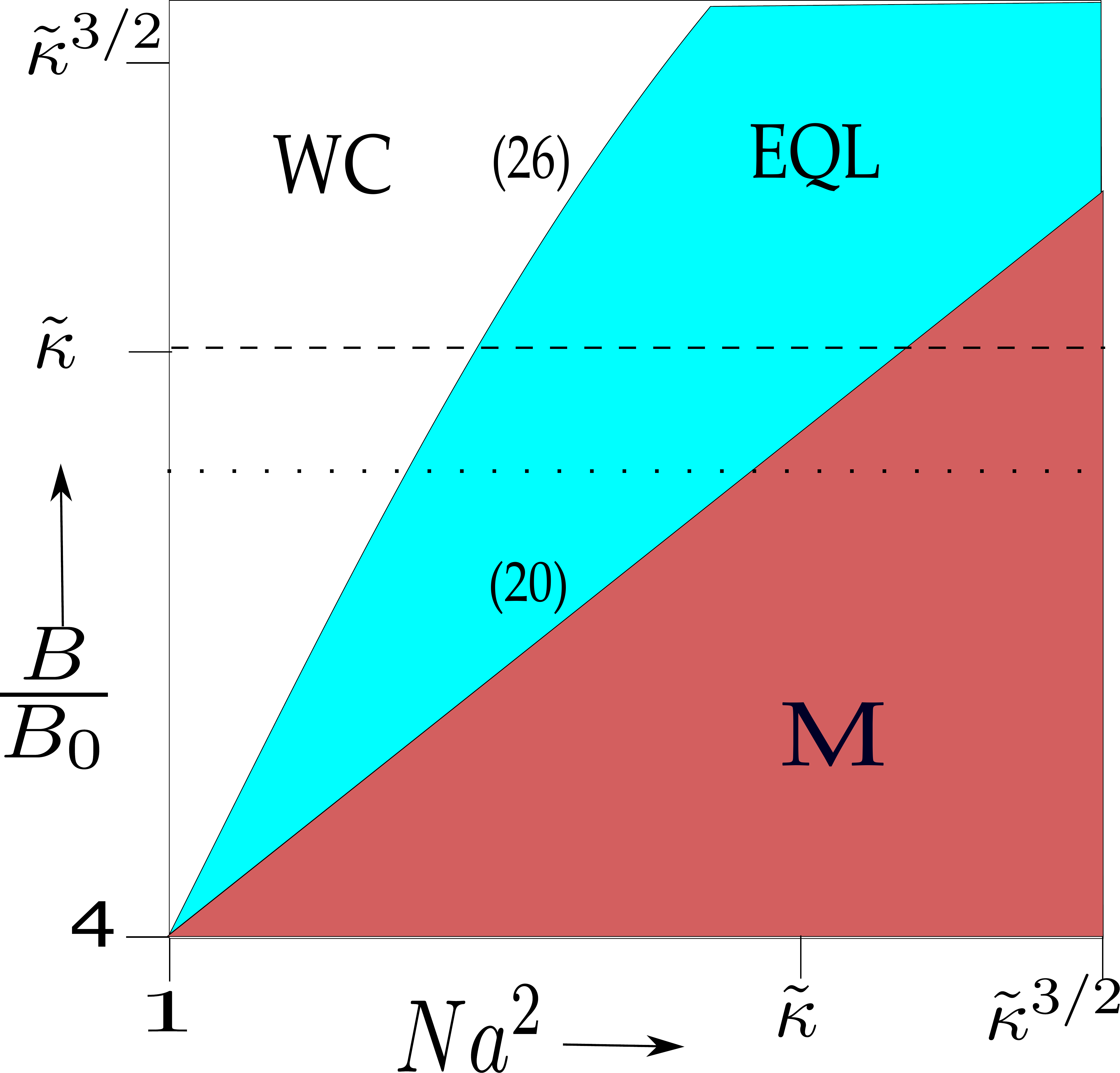}
	\caption{Phase diagram of the electron gas for $\boldsymbol{B}\perp \boldsymbol{E}$ in the dimensionless plane of $B/B_0$ and the surface concentration $Na^2$ plotted in a log-log scale.  The regions are the quasi-classical metal (M), the metallic EQL phase (EQL), and the insulating Wigner crystal state (WC). The dashed line indicates the ratio $B_{max}/B_0$ in STO, while the dotted line is the same quantity in InSb. See Eqs. (\ref{eq:bohr radius}), (\ref{eq:B0}), and (\ref{eq:kappabar}) in the text for the definitions of $a$, $B_0$, and $\tilde{\kappa}$.}
	\label{fig:phasediagram2}
\end{figure}

\begin{figure}[t]
	\centering
	\includegraphics[width=8.49cm, height=7.5cm]{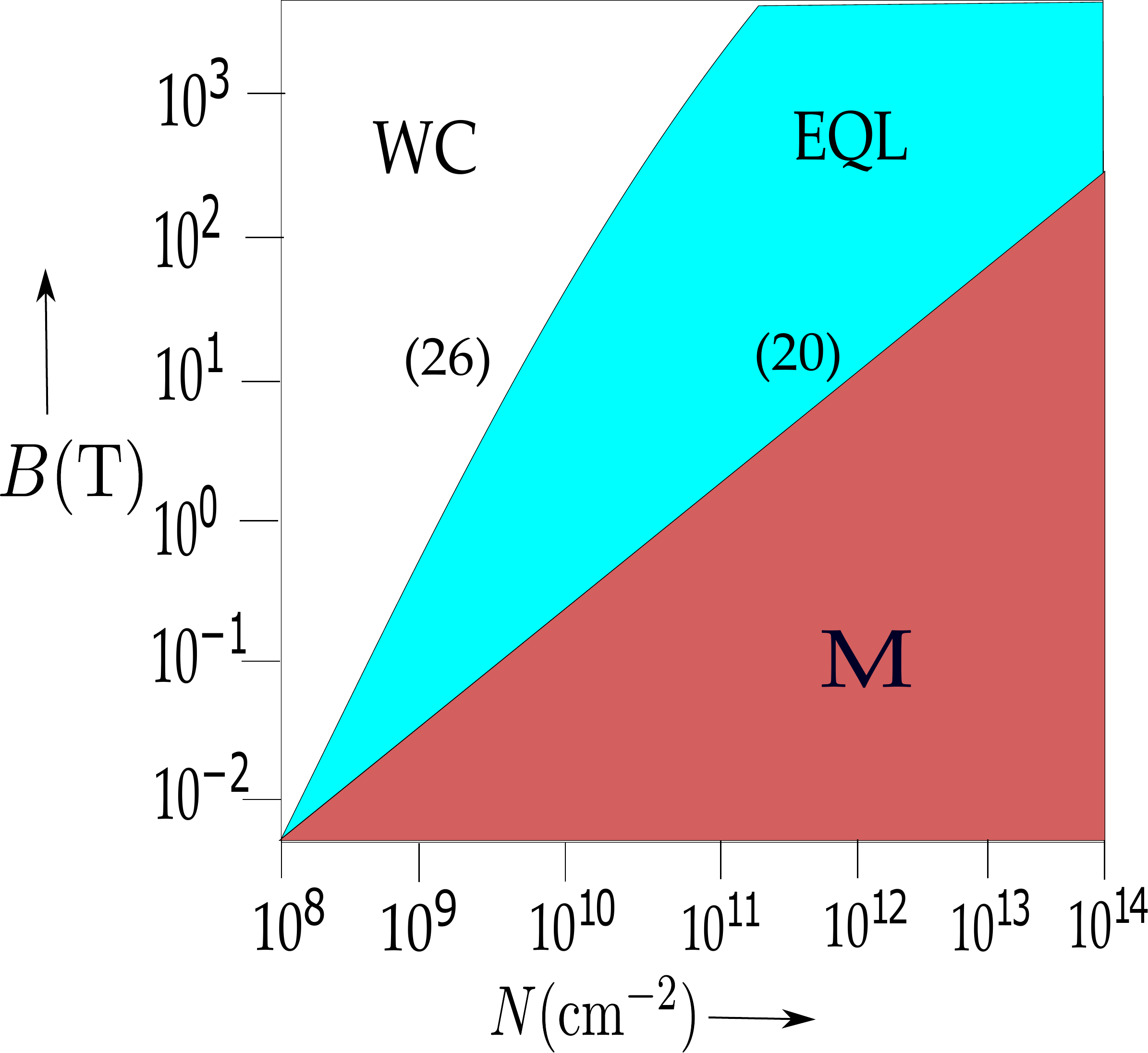}
	\caption{Phase diagram for $\boldsymbol{B}\perp \boldsymbol{E}$ in STO at liquid helium temperatures. The axes are $B(\text{T})$ and $N(\text{cm}^{-2})$ and the diagram is presented in a log-log scale. All regions and borders are identical to those in Fig. \ref{fig:phasediagram2}}
	\label{fig:phasediagramSTO}
\end{figure}

Within the EQL phase we calculate the Thomas-Fermi profiles of the electron density and electostatic potential as a function of the distance from the surface. By comparing the parameters found in the EQL metal with those of the quasi-classical metal (M), we determine the strength of the magnetic field at which the electrons enter the EQL. On the other end, these parameters are compared to those of the Wigner crystal (WC) phase at a large magnetic field to find the upper limit of the magnetic field at which the EQL metal is still valid. Fig. \ref{fig:phasediagram2} summarizes our results for the case of $\boldsymbol{B} \perp \boldsymbol{E}$ as a phase diagram in dimensionless units of $B/B_0$ and $Na^2$. Here
\begin{equation}
 B_0=m^{*2}e^3c/(\kappa^2\hbar^3)\label{eq:B0}
\end{equation}
  is the magnetic field such that $\lambda=a$. To preserve the universality of Fig. \ref{fig:phasediagram2} for different semiconductor parameters, we introduce the material specific constant 
  \begin{equation}
  \tilde{\kappa}=\frac{\kappa m_e}{m^*}.\label{eq:kappabar}
  \end{equation}
  In this notation, $B_0=(2.5\times10^{5})\: \tilde{\kappa}^{-2}$ T.

  Let us discuss what is achievable experimentally. The strongest static magnetic fields available in laboratories are approximately $B_{max}\simeq45$ T, from which it follows that 
  \begin{equation}
  B_{max}/B_0\approx1.8\times10^{-4}\:\tilde{\kappa}^2
  \end{equation}
  Given the values of $\kappa$ and $a$ in STO at liquid helium temperatures, we find $\tilde{\kappa}=1.3\times10^{4}$, and so $B_{max}/B_0\approx \tilde{\kappa}$, while for InSb, $\tilde{\kappa}=1.1\times10^{3}$ and $B_{max}/B_0\approx \tilde{\kappa}^{3/4}$. These values are indicated in Figs. \ref{fig:phasediagram2} and \ref{fig:phasediagram} by the dashed and dotted lines respectively. 
  
  Below we focus on STO. Fig. \ref{fig:phasediagramSTO} presents that phase diagram for $\boldsymbol{B} \perp \boldsymbol{E}$ in STO at liquid helium temperatures. 
  The lower EQL border defined by Eq. (\ref{eq:Bc1}) intersects $B_{max}$ in STO at a concentration $N\simeq 8\times 10^{12}$ cm$^{-2}$. Surface concentrations as low as $1\times10^{12}$ cm$^{-2}$ with high mobility have been achieved in modified LAO/STO interfaces and $\delta$-doped STO,\cite{TrierQHE,MatsubaraQHE} so that the lower critical magnetic field is reachable. Additionally, this range of surface concentrations $N=1\times10^{12}-8\times 10^{12}$ cm$^{-2}$ corresponds to bulk concentrations $N/d$ ranging between $3\times 10^{17}$ cm$^{-3}$ and $1\times 10^{18}$ cm$^{-3}$, where according to the data\cite{Spinelli} a reasonably large mobility can be maintained making the EQL achievable in this range of concentrations. Here $d$ is the characteristic width of the accumulation layer. 
  
The rest of the paper is organized as follows. In Sec. \ref{sec:accumulation} the new density profile $n(x)$ in the EQL is derived using the Thomas-Fermi approximation and the critical magnetic field at which the gas enters the EQL is found. In Sec. \ref{sec:phase} we finish constructing the phase diagram for different directions of magnetic field and arrive at Figs. \ref{fig:phasediagram2} and \ref{fig:phasediagram}. In Sec. \ref{sec:magnetocapacitance} we calculate the quantum capacitance in all phases and present the plot of the effective width of the inverse capacitance as a function of the magnetic field $B(\text{T})$ in Fig. \ref{fig:dq}.  In Sec. V we show how our results map to the problem of heavy atoms in ultrastrong magnetic fields which has been intensely studied in astrophysics. 
\section{Thomas-Fermi Theory of the Accumulation Layer}\label{sec:accumulation}
\subsection{Quasi-classical Metal}

In an accumulation layer, an electric field $E$ applied perpendicular to the the surface causes electrons to accumulate with a three-dimensional concentration $n(x)$, where $x$ is the distance measured from the surface. Here we assume that the semiconductor is such that the Fermi level in the bulk of the material lies at the bottom of the conduction band, and the electron concentration tends to zero at large distances. This can be true if the semiconductor is lightly doped by donors.
\footnote{Our theory of the accumulation layer is still valid if instead the semiconductor is intrinsic or even lightly doped by acceptors. If the semiconductor is intrinsic, then the bulk Fermi level lies in the middle of the gap. However, the system only acquires it's bulk value at a distance on the order of the screening radius of thermally excited electron-hole pairs, which is exponentially large at low temperatures, so that the effects of this can be neglected. In the case of light doping by acceptors our accumulation layer becomes an inversion layer which is followed by a large depletion layer of negatively charged acceptors. If the 3d concentration of acceptors is small, then it follows that the associated surface concentration of the depletion layer is small compared to our surface concentration $N$. In this case, the electric field is almost entirely screened by the inversion layer and again there is no difference from our theory.}
This problem was first solved in the absence of a magnetic field by Frenkel,\cite{Frenkel} and we repeat his argument below. 

In order to find the density profile, we make use of the Thomas-Fermi approach in which the local potential $\varphi(x)$ is related to the local chemical potential $\mu(x)$ such that $e\varphi(x)+\mu(x)=\varepsilon_F=0$.  In a normal metal, the chemical potential is related to the density such that

\begin{equation}
\mu_{_0}(x)=\frac{\hbar^2}{2m}[3\pi^2n_{_0}(x)]^{2/3}.\label{eq:chemical potential metal}
\end{equation} 
When the dielectric response is linear, the potential and density are related through Gauss's law, such that 
\begin{equation}
\frac{d^2\varphi_{\text{\tiny{$0$}}}}{dx^2}=\frac{4\pi e}{\kappa}n_{\text{\tiny{$0$}}}(x). \label{eq:gauss's law}
\end{equation} 
Combining Eqs. (\ref{eq:chemical potential metal}) and (\ref{eq:gauss's law}) with the above equilibrium condition, we obtain the Thomas-Fermi Equation
\begin{equation}
\frac{d^2}{dx^2}\left(\frac{\varphi_{\text{\tiny{$0$}}}}{e/a}\right)=\frac{2^{7/2}\kappa^{1/2}}{3\pi a^2}\left(\frac{\varphi_{\text{\tiny{$0$}}}}{e/a}\right)^{3/2}.\label{eq:Thomas Fermi Metal}
\end{equation}
The solution of this equation that satisfies the condition $\lim_{x\rightarrow\infty}\varphi(x)=0$ is known to be 
\begin{equation}
\varphi_{\text{\tiny{$0$}}}(x)=C_1\frac{e}{\kappa}\frac{a^3}{(x+d_{\text{\tiny{$0$}}})^4}\label{eq:potential metal},
\end{equation}
and the associated density is 
\begin{equation}
n_{\text{\tiny{$0$}}}(x)=C_2\frac{a^3}{(x+d_{\text{\tiny{$0$}}})^6}\label{eq:density metal}
\end{equation}
where $C_1=(225\pi^2/8)\simeq278$ and $C_2=(1125\pi/8)\simeq442$. 

To determine the characteristic width $d_{\text{\tiny{$0$}}}$, we use the definition of the two-dimensional electron density
\begin{equation}
N=\int_0^{\infty}n(x)dx.\label{eq:2d density}
\end{equation}
Combining Eqs. (\ref{eq:density metal}) and (\ref{eq:2d density}) we find that for the quasi-classical metal
\begin{equation}
d_{\text{\tiny{$0$}}}=C_3a\left(\frac{1}{Na^2 }\right)^{1/5}\label{eq:length in normal metal},
\end{equation}
where $C_3=\left(225\pi/8\right)^{1/5}\simeq2.45$.

\subsection{Extreme Quantum Limit}\label{sec:EQL}

The main purpose of this paper is to understand how the above distribution changes when the gas is subjected to such strong magnetic fields that it is in the EQL.

As stated above, when in the EQL, the kinetic energy in the direction perpendicular to the field is quantized and electrons occupy the lowest Landau level. This means that the density of electrons in the direction perpendicular to the field is fixed by the density of the lowest Landau level $1/(2\pi\lambda^2)$. In addition, the magnetic field aligns the spins in the direction of the field, lifting the spin degeneracy. The remaining direction has a density determined by the wavevector $k$. We can relate the maximum value of this wave vector to the three dimensional density of electrons by
\begin{equation}
\left(\frac{k}{\pi}\right)\frac{1}{2\pi\lambda^2}=n(x).\label{eq:momentum density in EQL}
\end{equation}
As a result, the local chemical potential changes from Eq. (\ref{eq:chemical potential metal}) to 
\begin{equation}
\mu(x)=\frac{\hbar^2}{2m}[2\pi^2\lambda^2n(x)]^2. \label{eq:chemical potential EQL}
\end{equation}
Proceeding in the same way as before, we arrive at the EQL Thomas-Fermi equation

\begin{equation}
\frac{d^2\varphi}{dx^2}=\frac{2^{3/2}e^{1/2}}{\pi(\lambda^4\kappa a)^{1/2}}\varphi^{1/2}. \label{eq:TF EQL}
\end{equation}
The solution gives the potential

\begin{equation}
\varphi(x)=C_4 \frac{e}{\kappa a}\frac{(x-d_{_\lambda})^4}{\lambda^4}\label{eq:potential EQL},
\end{equation}
and density as
\begin{equation}
n(x)=C_5 \frac{(x-d_{_\lambda})^2}{a\lambda^4}\label{eq:density EQL},
\end{equation}
where $C_4=1/(18\pi^2)\simeq0.006$ and $C_5=1/(6\pi^3)\simeq0.005$.

Using Eq. (\ref{eq:2d density}), the characteristic width is determined to be

\begin{equation}
d_{_\lambda}=C_6(Na\lambda^4)^{1/3},\label{eq:length in EQL}
\end{equation}
where $C_6=(18\pi^3)^{1/3}\simeq8.23$.

This result is valid when $d_{_\lambda}<d_0$. We find then that the magnetic field compresses the accumulation layer closer to the surface. Equating Eqs. (\ref{eq:length in normal metal}) and (\ref{eq:length in EQL}), and going back to the magnetic field, we find that the EQL is achieved when
\begin{equation}
B>B_{c1}=C_7B_0 (Na^2)^{4/5}\label{eq:Bc1},
\end{equation}
where $C_7=(C_6/C_3)^{3/2}\approx6.15$.
If $B_0<B<B_{c1}$, $n(x)$ obeys Eq. (\ref{eq:density metal}) until $\mu(x)=\hbar\omega_c$ where the gas enters the EQL. The distance from the surface at which this occurs is given by 
\begin{equation}
x_{\text{\tiny{$\lambda$}}}\approx(\lambda a)^{1/2}.
\end{equation}
 At this distance, the remaining electrons are in the EQL and the electron density is sharply cut off.  
  
We emphasize that the direction of the magnetic field has played no role in our discussion so far. Therefore, we see that our boundary given by Eq. (\ref{eq:Bc1}) is independent of the field direction. This line is shown in both Fig. \ref{fig:phasediagram2} and Fig. \ref{fig:phasediagram}. For $B\gg B_{c1}$, these diagrams lose their universality and we discuss them separately in the following section.

\section{Phase Diagrams for different Magnetic field directions}\label{sec:phase}

\begin{figure}[t]	
	\includegraphics[width=7.19cm, height=7.0cm]{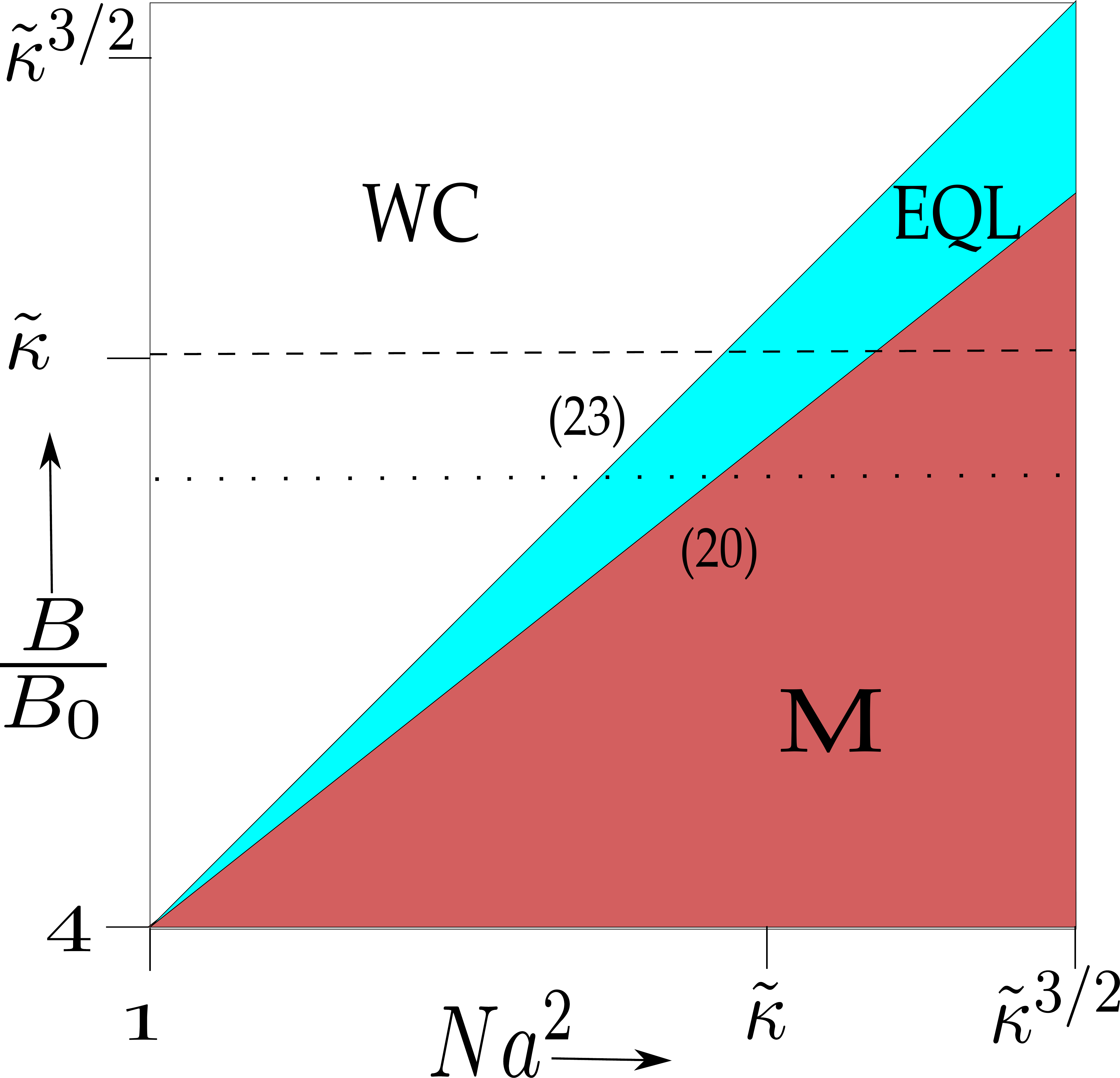}
	\caption{Phase diagram of the electron gas for $\boldsymbol{B}\parallel \boldsymbol{E}$ in the dimensionless plane of $B/B_0$ and the surface concentration $Na^2$ plotted in a log-log scale.  The regions are the quasi-classical metal (M), the metallic EQL phase (EQL), and the insulating Wigner crystal state (WC). The dashed line indicates the ratio $B_{max}/B_0$ in STO, while the dotted line is the same quantity in InSb. See Eqs. (\ref{eq:bohr radius}), (\ref{eq:B0}), and (\ref{eq:kappabar}) in the text for the definitions of $a$, $B_0$, and $\tilde{\kappa}$. Numerical values of $N$ and $B$ for STO at liquid helium temperatures can easily be recovered from comparison of Figs. 3 and 4.}
	\label{fig:phasediagram}
\end{figure}

 Below we address the role of the magnetic field direction and complete the phase diagrams Figs. \ref{fig:phasediagram2} and \ref{fig:phasediagram}. Let us assume that the electric field is strong enough such that $Na^2>1$.
The Thomas-Fermi approximation is only valid as long as the electrons can be treated semiclassically. We can make this condition quantitative by requiring that 
\begin{equation}
\frac{1}{\pi}\int_0^{\infty}k_x(x)dx>1
\end{equation}
which is a generalization of the 1d particle in a box.
 This condition depends on the direction of the magnetic field relative to the electric field, and so below we consider separately the two cases $\boldsymbol{B}\parallel \boldsymbol{E}$ and $\boldsymbol{B}\perp \boldsymbol{E}$.

\subsection{Magnetic Field Parallel to Electric Field}

When $\boldsymbol{B}\parallel \boldsymbol{E}$, $k_x(x)=2\pi^2\lambda^2n(x)$. From this we find that the approximation breaks down when
\begin{equation}
B=B_{c2}=B_0(2\pi Na^2)\label{eq:Bc2}.
\end{equation}
This is the boundary between regions EQL and WC in Fig. \ref{fig:phasediagram}. As the magnetic field is increased beyond this value, the Thomas-Fermi approximation becomes invalid everywhere. Instead, the electron gas forms a WC consisting of single electron cylinders of radius $\lambda$ and height $L$ (see Fig. \ref{fig:WCparallel}). The height of the cylinders can be determined as follows. At the border $B_{c2}$, the electron gas is confined to the first sub-band of a triangular potential well. The kinetic energy is then 
\begin{equation}
K=\frac{\hbar^2}{2mL^2}\label{eq:kinetic energy}
\end{equation}
while its potential energy is $U=eEL/2$, where $E$ is related to $N$ by Eq. (\ref{eq:Surface Charge}). Equating the kinetic and potential energies we find that 
\begin{equation}
L=(a/2\pi N)^{1/3}.\label{eq:WC height linear}
\end{equation}
The height of the cylinders $L$ should agree parametrically with the size of the accumulation layer $d_{_\lambda}$ along the EQL-WC phase boundary Eq. (\ref{eq:Bc2}). Let us confirm this.  Along the boundary, we know that Eq. (\ref{eq:Bc2}) gives $N=1/\lambda^2$, so that $L=(\lambda^2a)^{1/3}$.  If instead we are coming from the EQL region, we use Eq. (\ref{eq:length in EQL}) and find that $d_{_\lambda}=(\lambda^2a)^{1/3}$. 

Eq. (\ref{eq:WC height linear}) is the same as the width of the first sub-band wave functions obtained for an inversion layer in an electric field $E$.\cite{ SternHoward,Stern} However, contrary to the inversion layer where electrons are delocalized in the plane perpendicular to the field $E$, electrons here are strongly localized by the magnetic field in a cylinder of size $\lambda$. This is also the simplest case of quantum screening.\cite{Horing,Shklovskii1973,Kosarev}

\subsection{Magnetic Field Perpendicular to Electric Field}
 
If $\boldsymbol{B}\perp \boldsymbol{E}$, $k_x=1/\lambda$. As a result, we find instead of Eq. (\ref{eq:Bc2}) that the Thomas-Fermi approximation fails when $B>B_0(Na^2)^2$. 
 We show below that the EQL phase forms a WC at a somewhat smaller field 
\begin{equation}
B_{c3}\approx\frac{B_0(Na^2)^2}{(\ln(Na^2))^2}.\label{eq:Bc5}
\end{equation} 
This is the boundary given in Fig. \ref{fig:phasediagram2}.

\begin{figure}[t]
	\subfloat{(a)\label{fig:WCparallel}
		\includegraphics[width=7.5cm, height=4cm]{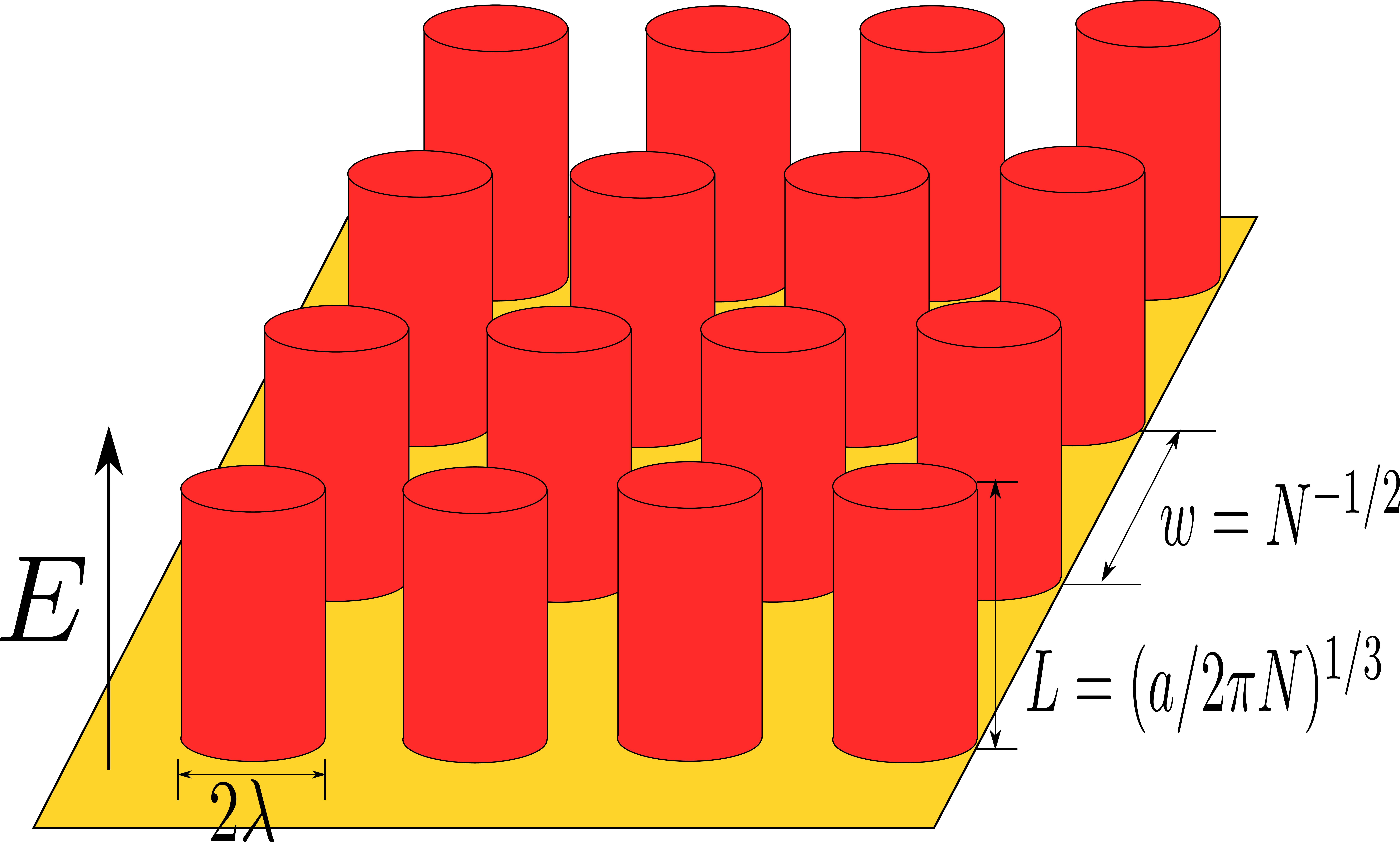}	
	}

	\subfloat{(b)\label{fig:WCperp}
		\includegraphics[width=8.5cm, height=4.8cm]{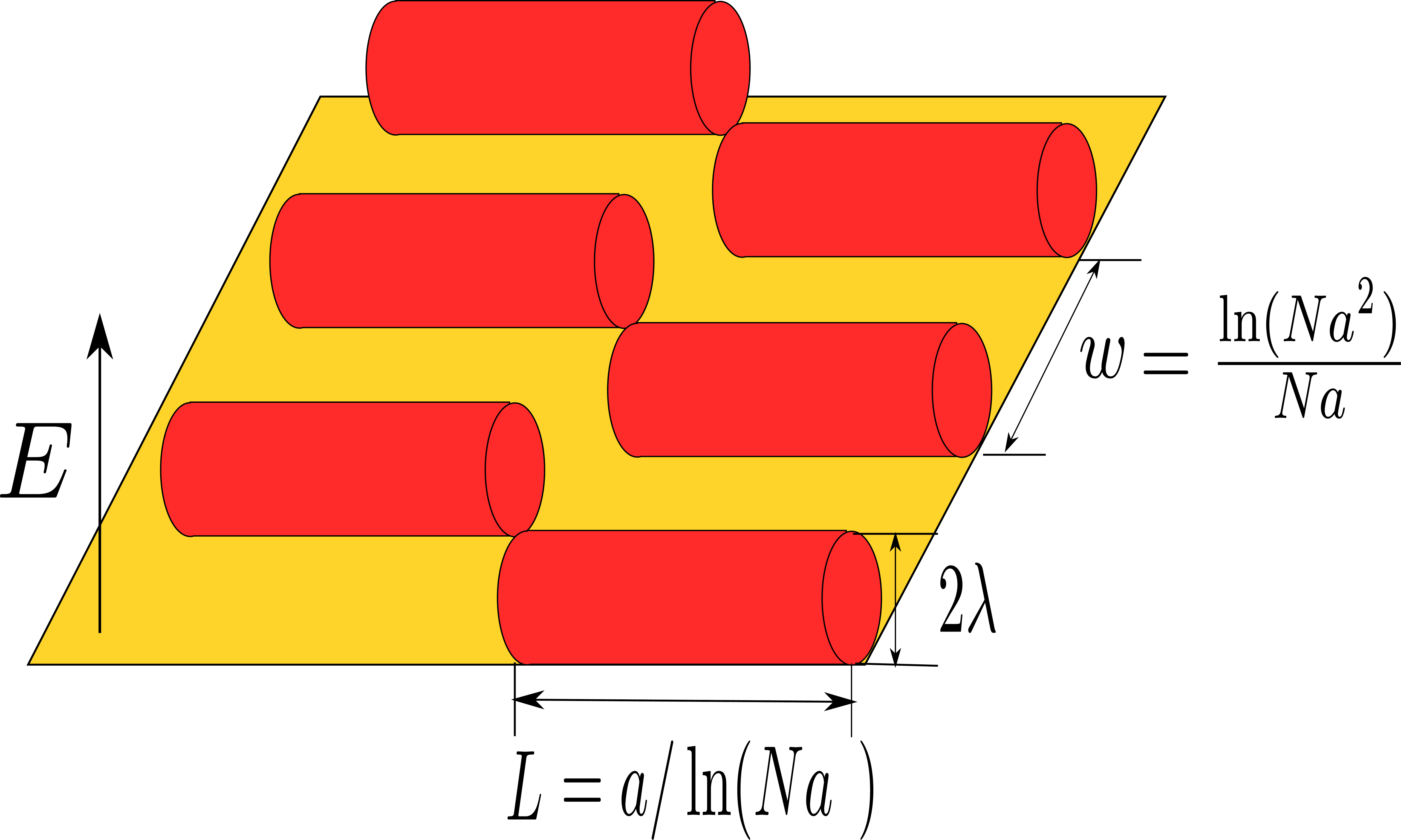}
	}
	
	\caption{(Color Online) Schematic of the electron structure in the WC phase for a)$\boldsymbol{B}\parallel\boldsymbol{E}$ and b)$\boldsymbol{B}\perp\boldsymbol{E}$. Each electron (red/dark grey) forms a cylinder of radius $\lambda$ oriented along the direction of the magnetic field  on the surface (yellow/light grey) inside the semiconductor. }
\end{figure}

The structure of the WC phase for $\boldsymbol{B}\perp \boldsymbol{E}$ is markedly different than when $\boldsymbol{B}\parallel \boldsymbol{E}$. We can imagine the electrons as cylinders of radius $\lambda$ oriented along $B$ which lie on their sides in the plane of the surface (See Fig. \ref{fig:WCperp}).

To describe the WC, one can imagine that $E$ is replaced by a uniform positive surface charge density $eN$ which is partitioned into Wigner-Seitz (WS) cells with charge $e$, length $L$, and width $w=1/NL$ so that each cell contains exactly one electron. We assume that the energy of each WS cell is approximately given by the sum of the kinetic energy Eq. (\ref{eq:kinetic energy}), and the electrostatic energy $U=-(e^2/\kappa L)\ln(NL^2)$. Optimization of this energy with respect to $L$ gives
\begin{equation}
L\simeq \frac{a}{\ln(Na^2)}.\label{eq:WCheight 2}
\end{equation}
As the magnetic field is reduced, it is natural to assume that the WC-EQL transition occurs when the electron is the same size as the WS cell.  Setting $w=\lambda$, and using Eq. (\ref{eq:WCheight 2}), we arrive at the border Eq. (\ref{eq:Bc5}). We see that the logarithmic term in the denominator of Eq. (\ref{eq:Bc5}) resembles those obtained previously for the metal-insulator transition in the bulk of a doped semiconductor in a strong magnetic field.\cite{Murzin1d, Shklovskii}

Up until now our theory is generic and is valid for any semiconductor material with a linear dielectric constant. In STO, however, the dielectric response becomes nonlinear at sufficiently high surface concentrations.\cite{KostyaSTO} It was shown the dielectric response becomes nonlinear when
\begin{equation}
Na^2=N_{c1}a^2=\frac{1}{\sqrt{\kappa}}\left(\frac{a}{a_{_0}}\right)^2\approx\tilde{\kappa}^{3/2}.
\end{equation}
Here $a_{_0}\simeq 3.9$ $\AA$ is the lattice constant in STO.
We see in Figs. \ref{fig:phasediagram2} and \ref{fig:phasediagram} that at this concentration, $B_{max}$ is such that the gas is still in region M, where the magnetic field only acts to cut the tail of the distribution. Thus, the EQL phase is unachievable experimentally when the dielectric response is nonlinear and so we limit Figs. \ref{fig:phasediagram2} and \ref{fig:phasediagram} to $Na^2<\tilde{\kappa}^{3/2}$. 

 \section{Magnetocapacitance}\label{sec:magnetocapacitance}

 In this section we calculate the capacitance of an accumulation layer as a function of the magnetic field for all phases. Our results can be used as tools for an experimental study of Figs. \ref{fig:phasediagram2} and \ref{fig:phasediagram}. For the setup we imagine that the accumulation layer is either created by the electric field of a metallic gate, or by a built in electric field $E$ to which a metallic gate adds a relatively small field $E$'. Examples of such devices include the gating of an intrinsic semiconductor by an ionic liquid and the application of a metallic gate to the top LAO surface in the LAO/STO heterostructure. In both cases one can study the differential capacitance per unit area $C=d(eN)/dV$, where $V$ is the gate voltage.  The inverse capacitance $C^{-1}$ may be written as the sum of the inverse geometrical capacitance  and the inverse quantum capacitance
 \begin{equation}
  C_q^{-1}=\frac{4\pi d_q}{\kappa}. \label{eq:quantum capacitance}
  \end{equation}
   Below we calculate $d_q$ for our phases M, EQL, and WC in both $\boldsymbol{B}\parallel \boldsymbol{E}$ and $\boldsymbol{B}\perp \boldsymbol{E}$ cases.

Let us discuss some of these results. In Sec. II the Thomas-Fermi potential profiles were found for the metallic M and EQL phases.  From Eq. (\ref{eq:potential metal}), we find that in region M the potential difference from $x=0$ to $x=\infty$ at a given concentration is 
 
 \begin{equation}
 \varphi(N)=\pi\left(\frac{225\pi}{8}\right)^{1/5}\frac{e}{\kappa a}(Na^2)^{4/5}.
 \end{equation}

Taking the derivative $d\varphi/dN$ and using Eq. (\ref{eq:quantum capacitance}), one finds
 
 \begin{equation}
 d_q(M)=\frac{d_0}{5}.\label{eq:dqM}
 \end{equation}
 
 In the EQL we know that the potential is instead given by Eq. (\ref{eq:potential EQL}) where $d_{\lambda}$ is given by Eq. (\ref{eq:length in EQL}). Proceeding in the same way, we find that the EQL changes the capacitance to 
 
 \begin{equation}
 d_q(EQL)=\frac{d_\lambda}{3}\label{eq:dqEQL}.
 \end{equation}
 Eqs. (\ref{eq:dqM}) and (\ref{eq:dqEQL}) are not surprising. In both cases, $d_q$ is the width of the accumulation layer in the direction of the electric field $E$, up to some numerical prefactor. To put another way, the effective width of the quantum capacitor is the width of the accumulation layer.  
 
 In order to find the point at which the capacitance transitions from that of the quasi-classical metal to the EQL metal, we equate Eqs. (\ref{eq:dqM}) and (\ref{eq:dqEQL}). We find then that the EQL becomes observable in capacitance measurements at $B_{c1}'=(5/3)^{3/2}B_{c1}$, which is slightly larger than Eq. (\ref{eq:Bc1}). At this field we should see the effects of the EQL begin to emerge, and so we use this as the field at which the gas transitions. 
 
Eqs. (\ref{eq:dqM}) and (\ref{eq:dqEQL}) are valid for both $\boldsymbol{B}\parallel \boldsymbol{E}$ and $\boldsymbol{B}\perp \boldsymbol{E}$ cases. As the magnetic field is increased, the two cases separate because of their different WC structure. We first discuss the $\boldsymbol{B}\parallel \boldsymbol{E}$ case.

 \begin{figure}[t]
 	\includegraphics[width=8.5cm, height=6.5cm]{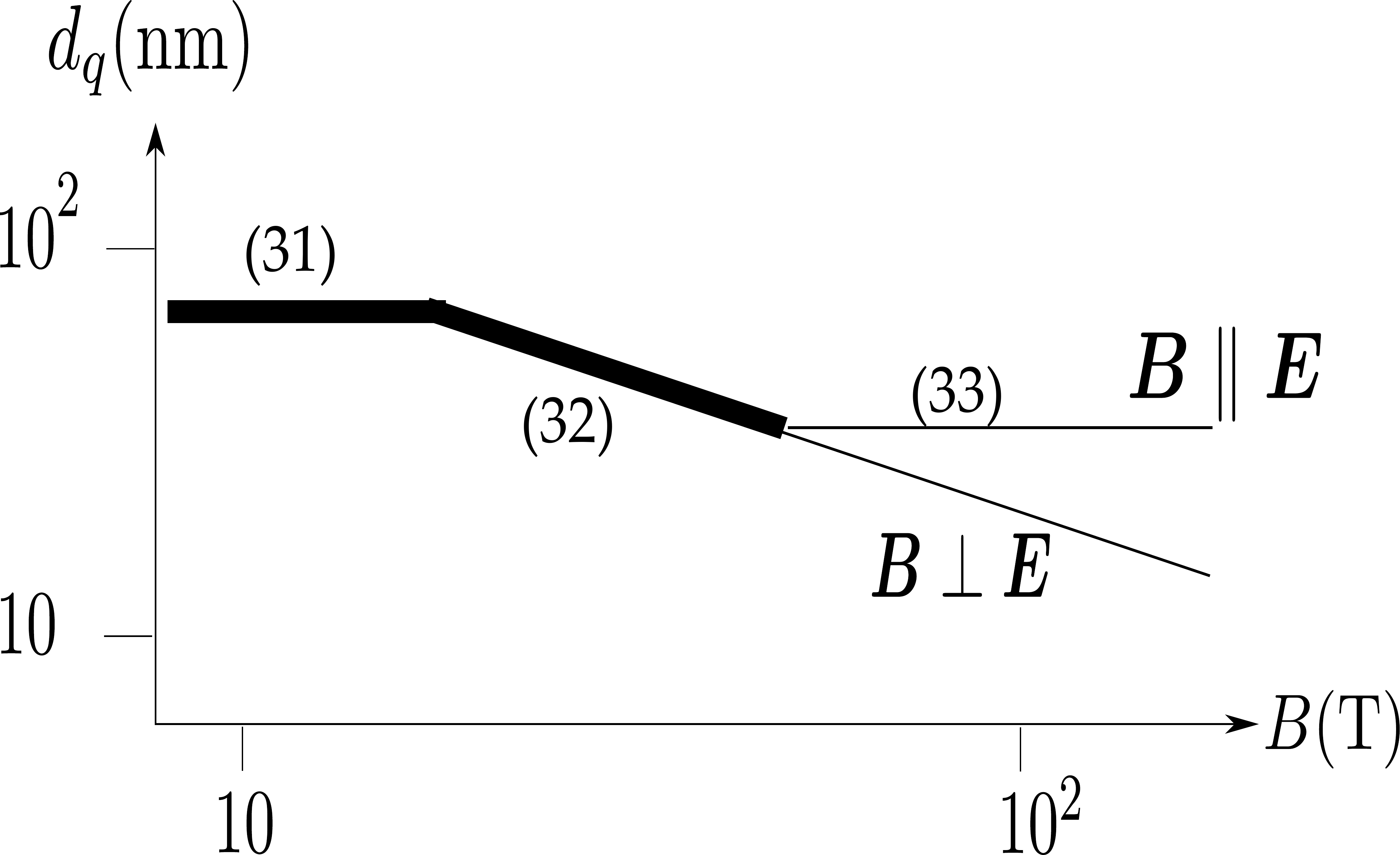}
 	\caption{Log-log plot of $d_q(\text{nm})$ as a function of $B(\text{T})$ for the phases of Figs. \ref{fig:phasediagram2} and \ref{fig:phasediagram} in STO at liquid helium temperatures with a surface concentration $N=10^{12}$ cm$^{-2}$. The region in which $d_q$ is the same for both $\boldsymbol{B}\parallel \boldsymbol{E}$ and $\boldsymbol{B}\perp \boldsymbol{E}$ is illustrated by a thick line. The numbers in parentheses correspond to equations in the text. }
 	\label{fig:dq}
 \end{figure}
As the magnetic field is increased, we see from Eq. (\ref{eq:dqEQL}) that $d_q$ will decrease. When $B=B_{c2}$ (Eq. (\ref{eq:Bc2})), a Wigner transition occurs and the gas enters the WC region. In this state, the width of the accumulation layer in the direction of $E$ is approximately given by Eq. (\ref{eq:WC height linear}) and no longer depends on the magnetic field. In order to find the value of the capacitance, we combine Eqs. (\ref{eq:Bc2}) and (\ref{eq:dqEQL}) and find
\begin{equation}
d_q(\text{WC}_{\parallel})=C_8 \frac{a}{(Na^2)^{1/3}} \label{eq:dqWC parallel}
\end{equation}
where $C_8=(\pi/6)^{1/3}\approx0.81$.

In the $\boldsymbol{B}\perp \boldsymbol{E}$ case, the transition to the WC phase happens at a much larger $B$ given by Eq. (\ref{eq:Bc5}). At this value of the field, the width of the accumulation layer is such that $d_q \sim 1/Na$.
If the field is increased further, then $d_q$ continues to decrease as the negative energy due to correlation effects of the WC begin to dominate.\cite{Bello} However at such large fields, the distance between electrons may become comparable to the distance between the WC and the gate, and the coupling of electrons to their image charge becomes the dominant factor in the determination of the capacitance.\cite{SkinnerCap} Despite this, such magnetic fields are too high to reach experimentally, and so we refrain from any further discussion of this limit.

We summarize these results in Fig. \ref{fig:dq} as a plot of $d_q(\text{nm})$ vs. $B(\text{T})$ for STO samples with a surface concentration $N=10^{12}$ cm$^{-2}$. At this concentration, the transition into the EQL occurs when $B=B_{c1}'\approx 18$ T while the EQL-WC transition occurs at $B=B_{c2}\approx 46$ T. We see then that the EQL phase is within the realistic range of magnetic fields and so capacitance measurements at this concentration provide an opportunity in which the first border will be observed. However if one wishes to see the splitting between the two directions, one needs to go to lower concentrations than $N=10^{12}$ cm$^{-2}$.
\section{Heavy Atoms in Pulsars}
So far, we have restricted our discussion to electron accumulation layers in semiconductor materials with such large Bohr radii that the EQL is achievable experimentally. For atoms, the EQL is achieved when the magnetic field is larger than $B_0=2.34\times 10^5$ T and is completely unattainable in a laboratory setting. However, in rotating neutron stars, or pulsars, the magnetic fields at the surface range from $10^8-10^9$ T,\cite{Gold} so that the EQL can be achievable even for atoms. The effect of the large magnetic field on the structure of the surface layer of neutron stars has been studied extensively.\cite{LibermanMatter}  

It is believed that within the surface of neutron stars there exists a layer enriched by iron atoms.\cite{Shapiro} Motivated by this, Kadomtsev studied heavy atoms in ultrastrong magnetic fields, where he used an EQL Thomas-Fermi equation which is the spherically symmetric analog of our Eq. (\ref{eq:TF EQL}).\cite{Kadomtsev,Kudryavtsev} He found that the EQL Thomas-Fermi description of the atom is valid as long as the magnetic field is in the range
\begin{equation}
Z^{4/3}\ll\frac{B}{B_0}\ll Z^3 \label{eq:Kadomtsevs border}
\end{equation} 
where $Z$ is the nuclear charge of the atom. When $B/B_0\ll Z^{4/3}$, the magnetic field has only a perturbative effect on the atomic structure, while for $B/B_0\gg Z^3$, the atom is elongated along the direction of the magnetic field. We will now show that there exists a mapping between the nuclear charge $Z$ and the surface concentration $N$ showing that the $Z^{4/3}$ and $Z^3$ borders are in agreement with those we found for the EQL phase when $\boldsymbol{B}\parallel\boldsymbol{E}$.

We can imagine that within the accumulation layer, electrons are bound at a distance $d$ away from the surface by the positively charged plane with charge density $eN$. We can think that this plane consists of positive squares (nuclei) of length $d$ and charge 
\begin{equation}
Z=Nd^2. \label{eq:nuclear charge}
\end{equation}
At the lower critical field $B_{c1}$, we know that the characteristic width of the gas is given by Eq. (\ref{eq:length in normal metal}). Using Eq. (\ref{eq:nuclear charge}) we find that the nuclear charge at this field is $Z=(Na^2)^{3/5}$.  From here, we find then that the lower critical field $B_{c1}$ given by Eq. (\ref{eq:Bc1}) is related to the nuclear charge by 
\begin{equation}
\frac{B_{c1}}{B_0}=(Na^2)^{4/5}=Z^{4/3}
\end{equation} 
in agreement with the lower border of Eq. (\ref{eq:Kadomtsevs border}). On the other hand, we know that if the magnetic field is applied parallel to the electric field, the Thomas-Fermi approximation fails when $B=B_{c2}$, where $B_{c2}$ is given by Eq. (\ref{eq:Bc2}). At this field strength, the width of the layer $d$ is given by Eq. (\ref{eq:length in EQL}), from which it follows that  $Nd^2=(Na^2)^{5/3}(B_0/B)^{4/3}$. Solving this equation for $B$, and using Eq. (\ref{eq:Bc2}), we find that the EQL region ends when 
\begin{equation}
\frac{B}{B_0}=Z^3
\end{equation}
in agreement with Kadomtsev's second border. 

Note that in our comparison, we have used the case of $\boldsymbol{B}\parallel \boldsymbol{E}$. In a heavy atom, the magnetic field can be both parallel and perpendicular to the electric field of the nucleus. However, we know from Sec. III that when the $\boldsymbol{B}\parallel \boldsymbol{E}$, the Thomas-Fermi approximation fails at a smaller $B$ than for the case of $\boldsymbol{B}\perp \boldsymbol{E}$. It is for this reason that the mapping from the accumulation layer to the heavy atoms needs $\boldsymbol{B}\parallel \boldsymbol{E}$. 

Let us conclude with a discussion about the structure of the atom when $B/B_0\gg  Z^3$. At such fields all electrons are in the lowest Landau level and occupy a single sub-band in the direction of the field $B$. From the above mapping, it would seem natural to expect the structure to be similar to a WC where the same limits apply. Actually, due to the strength of the Coulomb field of the point charge $Z$, the electrons instead compress into a single uniformly charged cylinder of radius $R=\lambda Z^{1/2}$ and height $L=a/Z\ln(B/Z^3)$. The compression of the cylinder is stopped by the kinetic energy $\hbar^2/(2mL^2)$ along the direction of $B$. One can think that in our Fig. \ref{fig:phasediagram}, the atom becomes ``frozen" at the EQL-WC border when $B/B_0\gg Z^3$.

\section{Conclusion}
In this paper, electron accumulation layers induced by an electric field perpendicular to the surface are studied for semiconductors with a large Bohr radius under the influence of a very strong magnetic field.
Phase diagrams in the plane of magnetic field strength and two-dimensional electron concentration are found for two orientations of magnetic field with respect to the electric field (Figs. 2, 3, and 4). Each diagram is found to have three phases: the quasiclassical metal (M), the extreme quantum limit (EQL) electron gas, and
the Wigner crystal (WC).
We showed that in the case of STO, with the largest known Bohr radius $a=700$ nm, all of these phases may be
reached in the available magnetic fields. We calculate the width of the accumulation layer for all of the phases and predict
how the quantum capacitance of the accumulation layer changes with the magnetic field when we cross from one phase to another.
This provides a tool with which the phase diagrams can be studied experimentally with the help of magneto-capacitance measurements.
In the future, it will be interesting to explore the transport properties of the extreme quantum limit of STO accumulation layers.

Above we assumed that the spectrum near the bottom of the conduction band in STO consists of a single isotropic band with an effective mass of $m^*\approx1.5$ $m_e$. In reality, the band structure of STO consists of three degenerate 3d orbitals of the Ti atoms which are split by the spin-orbit interaction and the tetragonal distortion of STO at low temperatures.\cite{Lin,Mattheiss} At concentrations $n < 10^{18}$ cm$^{-3}$, only the lowest band is occupied and Shubnikov-de Haas oscillations measurements show the band has an effective mass of $1.5-1.8$ $m_e$.\cite{Lin} It was mentioned above that with surface concentrations $N \sim 10^{12}$ cm$^{-2}$ 
one can explore the EQL with the available magnetic fields. Such $N$ correspond to bulk concentrations $n \sim  10^{17}$ cm$^{-3}  \ll 10^{18}$ cm$^{-3}$, where the approximation of an effective mass of $m^*\approx1.5$ $m_e$ is justified.

$\phantom{}$
\vspace*{2ex} \par \noindent
{\em Acknowledgments.}

We are grateful to Eugene Kolomeisky, Nini Pryds, and Brian Skinner for helpful discussions. This work was supported primarily by the National Science Foundation through the University of Minnesota MRSEC under Award No. DMR-1420013.

\bibliography{EQLref}

\end{document}